\documentclass[prl,aps,superscriptaddress,preprint]{revtex4}
\usepackage{graphics}
\usepackage{amsmath}
\usepackage{amsfonts}
\usepackage{amssymb}
\usepackage{graphicx}
\begin{document}
\title{NiS - An unusual self-doped, nearly compensated antiferromagnetic metal}
\author{S. K. Panda}
\affiliation{Centre for Advanced Materials, Indian Association for the Cultivation of Science, Jadavpur, Kolkata-700032, India}
\author{I. Dasgupta}
\affiliation{Centre for Advanced Materials, Indian Association for the Cultivation of Science, Jadavpur, Kolkata-700032, India}
\affiliation{Department of Solid State Physics, Indian Association for the Cultivation of Science, Jadavpur, Kolkata-700032, India}
\author{E. \c{S}a\c{s}{\i}o\u{g}lu}
\affiliation{Peter Gr\"{u}nberg Institut and Institute for
Advanced Simulation, Forschungszentrum J\"{u}lich and JARA, D-52425 J\"{u}lich, Germany}
\author{S. Bl\"{u}gel}
\affiliation{Peter Gr\"{u}nberg Institut and Institute for
Advanced Simulation, Forschungszentrum J\"{u}lich and JARA, D-52425 J\"{u}lich, Germany}
\author{D. D. Sarma}
\altaffiliation{Also at Jawaharlal Nehru Center for Advanced Scientific Research, Bangalore-560054, India and Department of Physics and Astronomy, Uppsala University, Uppsala, Sweden.}
\email{sarma@sscu.iisc.ernet.in}
\affiliation{Centre for Advanced Materials, Indian Association for the Cultivation of Science, Jadavpur, Kolkata-700032, India}
\affiliation{Solid State and Structural Chemistry Unit, Indian Institute of Science, Bangalore-560012, India}
\affiliation{Council of Scientific and Industrial Research-Network of Institutes for Solar Energy (CSIR-NISE), New Delhi, India}

\begin{abstract}
NiS, exhibiting a text-book example of a first-order transition with many unusual properties at low temperatures, has been variously described
in terms of conflicting descriptions of its ground state during the past
several decades. We calculate these physical properties within first-principle approaches based on the density functional theory and conclusively establish that all experimental
data can be understood in terms of a rather unusual ground state of NiS that is best described as a self-doped, nearly compensated, antiferromagnetic metal,
resolving the age-old controversy. We trace the origin of this novel ground state to the specific details of the crystal structure, band dispersions
and a sizable Coulomb interaction strength that is still sub-critical to drive the system in to an insulating state. We also show how the specific antiferromagnetic structure is a consequence of the less-discussed 90$^o$ and less than 90$^o$ superexchange interactions built in to such crystal structures.
\end{abstract}
\maketitle
The electronic structure of hexagonal NiS has remained a classic problem in the field of metal-insulator transition (MIT), attracting sustained interest spanning a period
of several decades~\cite{neutron,sparks1,sparks2,Mott1,neutron1,resistivity,optical74,mattheiss,fujitheory,photo1,photo2,usuda,xas}. Its first order phase transition between
a highly conducting ($\sim 10^{-5}$ $\Omega$-cm) Pauli paramagnetic state at temperatures (T) greater than the transition temperature $T_t\sim$ 265 K and
an antiferromagnetic, low conductivity ($\sim 10^{-3}$ $\Omega$-cm) state for $T<265$~K is a matter of discussion in textbooks on MIT~\cite{RMP,mottMIT,Japanesebook}
for many years.
While every conceivable relevant experiment has been carried out with care for this system, the nature and the origin of the low temperature (LT) phase
has remained inconclusive. While all investigations
agree that the LT phase is a simple A type antiferromagnet,
the essential difficulty in describing the LT phase arises from its unusual electronic/transport properties. The basic issue in this context is whether the LT phase of NiS, a metal or an insulator.
In spite of this being an apparently simple issue to settle, a host of experimental techniques employed to address the issue of metallicity in the LT phase, have failed to resolve the controversy.
For example, the low temperature resistivity~\cite{resistivity},
which in itself is often a good
indicator of the ground state being a metal or an insulator, is in the range of $\sim 10^{-3}$ $\Omega$-cm which can be described as that of a
very bad/dirty metal or a poor insulator (either because of a very small gap or because of shallow impurity levels),
this value of the resistivity being in the range of Mott's minimum metallic conductivity.
Early measurements of the
Hall resistivity of NiS established a carrier concentration of $\sim 10^{20}$-$10^{21}$/cm$^3$, which is smaller than in good metals, but too large to be
reconciled with an insulator.
This would suggest NiS
to be a low carrier density metal.
However, it is known that NiS invariably forms with Ni deficiency, leading to a chemical formula of Ni$_{1-x}$S.
Hall measurements for a series of Ni$_{1-x}$S with
different values of $x$ indicated that the carrier concentration is a nearly linear function of $x$ and, more importantly, the extrapolated carrier concentration
at $x=0$ is very
close to zero, suggesting that the stoichiometric compound is, in fact, an insulator~\cite{resistivity}. Similar confusion also exists in the interpretation of
the photoconductivity data
which shows~\cite{optical99} a gap-like structure with a pronounced minimum at $\sim$ 180 meV followed by a small rise with further decrease in the frequency
($\omega$).
While the small but finite
photoconductivity for $\omega\rightarrow$ 0 can be interpreted as the signature of the Drude-peak in a bad metal, the pronounced gap like structure has been used
to suggest~\cite{optical99} NiS to be an
insulator with a band gap of about 180 meV. In this latter interpretation, the Drude-like feature is thought to arise from doped carriers due to a finite
concentration of Ni vacancy.
Photoemission studies
with its direct measurement of states at and around the Fermi energy ought to be able to resolve the controversy conclusively. However, very similar photoemission
data have been used in two different publications to imply an insulating~\cite{photo1} and a metallic ground state~\cite{photo2}, therefore, not helping to resolve
the controversy.
Photoemission experiments, however do establish that an energy gap, if at all existent in NiS, is not larger than a couple of meV~\cite{photo2}.
This is in direct contrast to the suggestion of a gap of about 180 meV based on optical conductivity data~\cite{optical99}.
\par
In view of such extensive and high quality experimental information available on the
LT phase of NiS, the ground state of this system appears to be
over-determined, making it an ideal case for theoretical investigations.
There have been several theoretical attempts~\cite{anisimov,usuda,mattheiss,Mott1,kasowski,tyler,ovchinnikov}
to calculate the electronic structure of the LT phase of NiS.
Some of these works~\cite{mattheiss,usuda,tyler,ovchinnikov} claim NiS to be non-metallic,
while almost equal number of studies~\cite{Mott1,anisimov,kasowski} claim that NiS is metallic.
Curiously, none of these earlier studies made use of the availability of the vast experimental information on NiS.
In this report, we present the electronic structure of NiS as a function of $U$ within the local spin density functional
theory (LSDA) +U, comparing theoretical results with experimental ones and finally resolving
the several decade long controversies concerning the ground state of NiS, establishing it as a novel example of a dilute,
self-doped and, consequently nearly compensated p-d metal. In addition, we establish the origin of the specific magnetic structure found in the
LT phase of NiS, underscoring the importance of the less discussed super-exchange interaction through $\leq$ $90^{\circ}$ bond angle.
These mechanisms being dependent solely on the geometric structure, present considerations are of relevance not only to all compounds forming in
the NiAs structure, but also more generally, to any compound that contains face-sharing octahedral arrangement of the transition metal ions.
\section*{Results}
\begin{figure}
\includegraphics[width=\columnwidth]{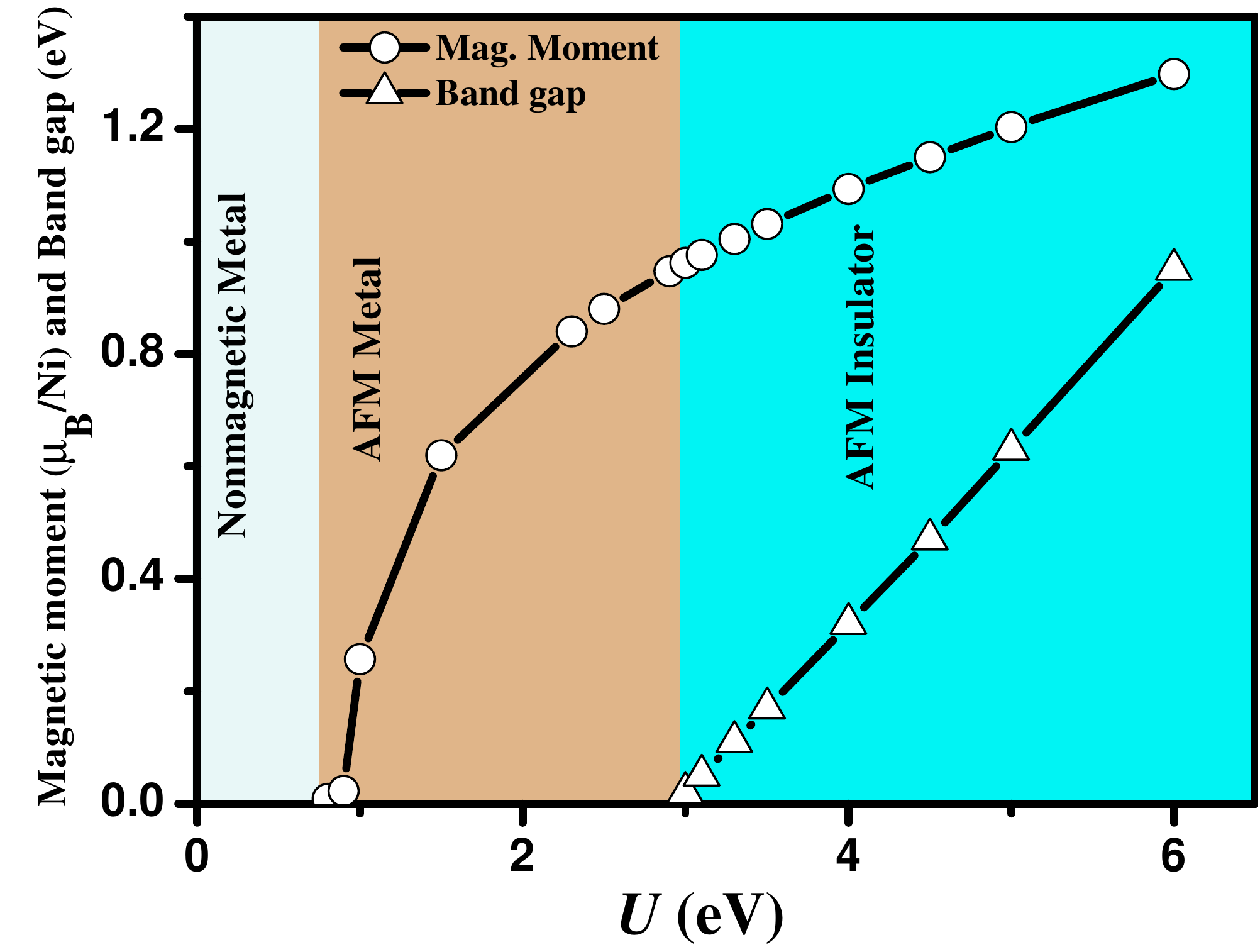}
\caption{(Color online) The phase diagram for the ground state of NiS as a function of $U$. Change of the magnetic moment per Ni site and the band gap in the antiferromagnetic phase
as a function of $U$ are also shown.}
\label{bandgap}
\end{figure}

In agreement with an earlier study~\cite{fujitheory}, our self-consistent FPLAPW spin-polarized calculations for both LT and
high temperature (HT) crystal structures
in the framework of LSDA do not yield an antiferromagnetic ground state and converged to non-magnetic metallic solutions.
Results of our LSDA+U calculations as a function of \textit{U} are shown in Fig.~\ref{bandgap}
in terms of the magnetic moment and the band gap. Thus, Fig.~\ref{bandgap} clearly establishes a phase diagram for the ground state of NiS,
showing that the ground state is a non-magnetic
metal for $U<0.8$~eV, an antiferromagnetic metal 0.8~eV~$\leq$~$U$~$<$~3.0~eV and an antiferromagnetic insulator for \textit{U} $\geq$~3.0~eV.
Expectedly, the magnetic moment increases monotonously and the band gap nearly linearly
with $U$, beyond two critical values of $U$, $U_{c1} \simeq$ 0.8 eV and $U_{c2} \simeq$ 3.0 eV for the magnetic transition and
the metal-insulator transition, respectively.
We do not stress the exact numerical values of the two $U_c$'s, since it is known that the value of \textit{U} required to reproduce
experimental results within LSDA+U is in general smaller than the \textit{U} estimated from constrained density functional calculations~\cite{NiO_prb2000}.
However we believe that the qualitative features of the phase diagram is robust.
\par
Ground state antiferromagnetic structures for both the metallic and the insulating states are
 found to be the same A type antiferromagnet. This is the reason why
the magnetic moment does not show any anomaly across the metal-insulator boundary at $U_{c2}=3.0$ eV.
This particular magnetic structure can be understood on the basis of the filling of Ni-$d$ bands
and the geometric structure. In the presence of octahedral crystal field splitting in the NiAs structure, the t$_{2g}$ states are completely occupied, while the e$_g$ states
are half filled for Ni in the $d^8$ configuration in NiS.
Ni atoms along the (a,b) plane are coupled to each other with an $88^{\circ}$ Ni-S-Ni bond angle arising from the edge-sharing octahedra. This nearly
$90^{\circ}$ superexchange interaction via the S $2p$ states gives rise to a weak ferromagnetic coupling,
leading to the ferromagnetic arrangement of Ni atoms in the (a,b) plane.
The Ni atoms in the successive planes along the c-axis are, however, connected by the face-shared NiS$_6$ octahedra,
making Ni-S-Ni bond angle to be $68^{\circ}$. This allows for a substantial Ni $e_g$-S $2p$-Ni $e_g$ hopping, thereby making the Ni atoms in successive planes along the c-axis
coupled antiferromagnetically due to the superexchange interaction. We calculated the antiferromagnetic superexchange interaction to be quite strong (126.5 meV), based
on an effective Ni-$d$ model. This is consistent with the reported very high value of the NiS Neel temperature~\cite{neutron1}.
\par
As for the electronic structure, present  calculations indicate extensive mixing of Ni $d$ and S $p$ states for the range of values of $U$ (0.8~eV~$\leq$~$U$~$<$~3.0~eV) where we find the
antiferromagnetic metallic phase, as shown in the Supplemental Material (SM) in terms site and orbital projected partial densities of states and fat-band representation
of the energy dispersions. This establishes NiS within this range of $U$-values to be a $p$-$d$ metal. The presence of the nearby insulating phase with increasing $U$
confirms the correlated nature of this $p$-$d$ metallic phase close to the phase boundary. It is interesting to note that an antiferromagnetic metallic phase has indeed
been shown~\cite{ZSA_DD1,ZSA_DD2} to exist within the extended Zaanen-Sawatzky-Allen type phase diagram for such inorganic solids within the $p$-$d$ metallic phase adjacent to the covalent
insulator phase. However, the range of interesting properties of the LT phase of NiS, established experimentally, cannot be understood within such a simplistic generic description.
\par
\begin{figure}
\includegraphics[width=\columnwidth]{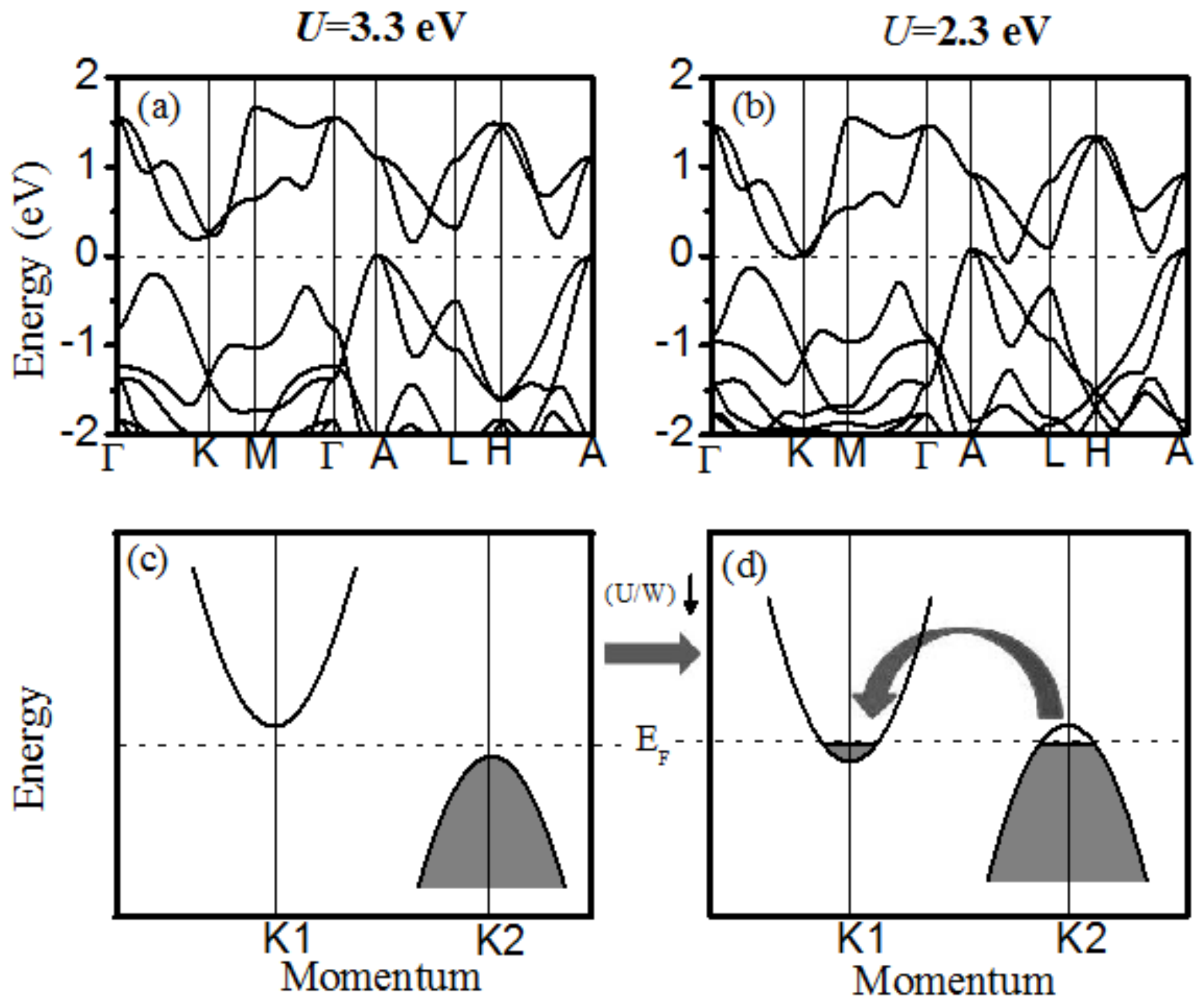}
\caption{Band dispersion along various high symmetry directions for (a) U = 3.3 eV and (b) 2.3 eV. Corresponding schematic diagrams are shown in (c) and (d).}
\label{bandstructure}
\end{figure}

We show the band dispersions along various high symmetry directions of the
Brillouin zone in Fig.~\ref{bandstructure}(a) and (b) for two values of $U$, namely 3.3 eV and 2.3 eV, corresponding to the antiferromagnetic insulating and metallic states,
respectively.
The essential points to note in these band dispersions are that (i) the insulating state has an indirect gap with the top of the valence band at the
A (0, 0, $\frac{1}{2}$) point and the
bottom of the conduction band at an intermediate k-point along the A-L direction; and (ii) the transition to the metallic state retains the band dispersions
almost identical to that of the insulating
state, with the conduction bands moving down in energy relative to the valence bands. These essential aspects
can be represented by the schematic band dispersions shown in Fig.~\ref{bandstructure}(c) and (d). From this schematic, it becomes obvious that the bandgap goes
continuously to zero  as $U$ approaches $U_{c2}$ from above, as also seen in our calculated results (Fig.~\ref{bandgap}).
As soon as $U$ begins to be lowered below $U_{c2}$, electrons from the top of the valence band are transferred to the bottom of the conduction band (and some other k-points depending on the $U$-value) of the Brillouin zone,
thereby a part of the $d$-band doping another part in the same $d$-manifold.
This self-doping ensures that the electron and the hole volumes contained by the Fermi surface are \textit{exactly} equal. This is not only true for the schematic dispersions
shown in Fig.~\ref{bandstructure}(d), but also for the full, all orbital calculation in Fig.~\ref{bandstructure}(b), independent of all details of the
full calculation such as the A point having two hole pockets
and the bottom of the conduction band having only one electron pocket as can be seen in the corresponding Fermi surface plots in the SM.
In this rather unexpected case of self-doping, the material may be in a nearly-compensated state, if the electron and hole mobilities are similar, as would be
suggested by similar curvatures of band dispersions defining the electron and hole pockets.
In the remaining part of this article, we explore the consequences of this unusual
ground state electronic structure {\it vis-a-vis} the main issue of insulating/metallic ground state of NiS.
\par
\begin{figure}
\includegraphics[width=\columnwidth]{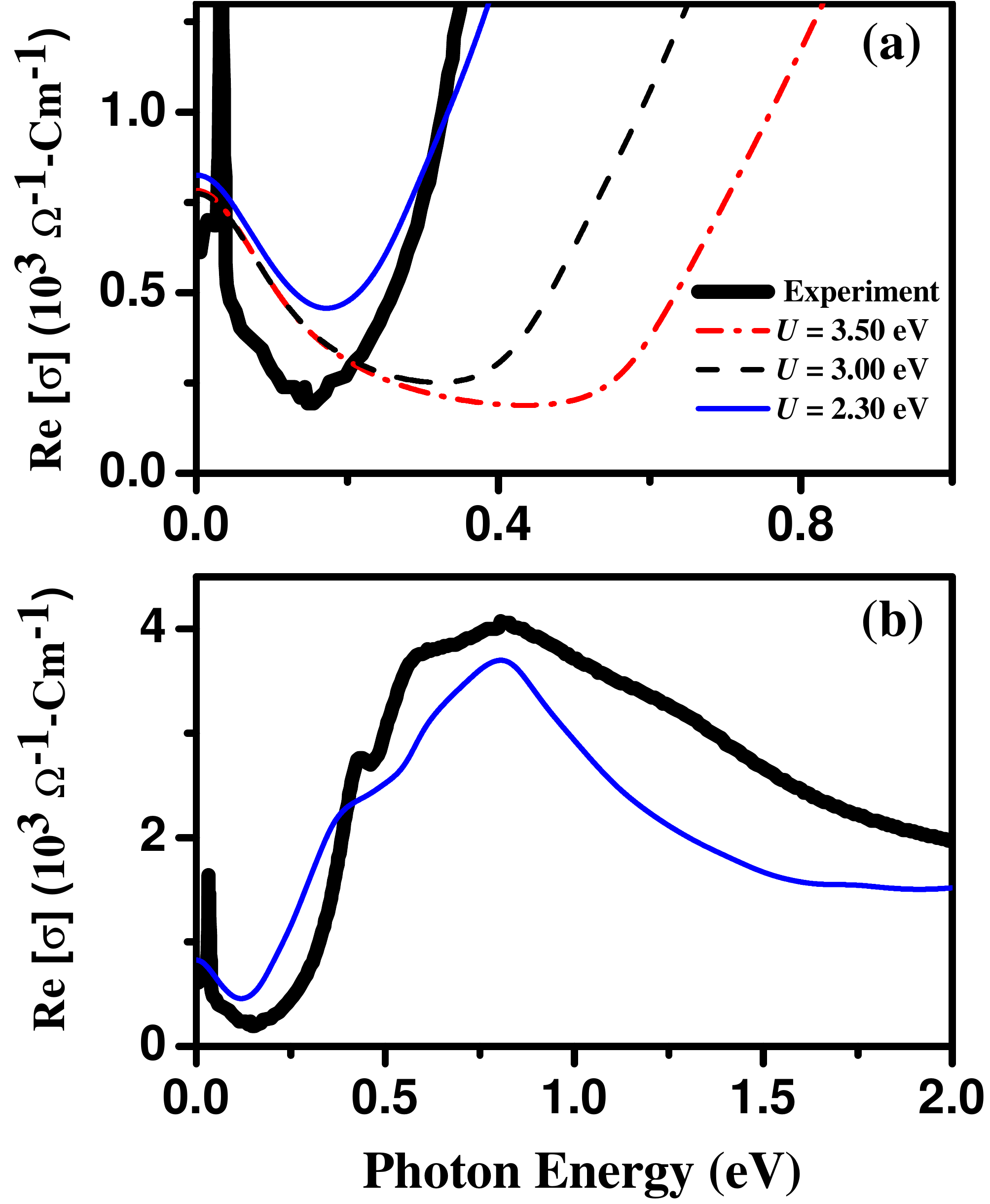}
\caption{Comparison of the experimental and theoretical optical conductivity of of Ni$_{.998}$S as discussed in the text. Experimental data are taken from Ref.~\onlinecite{optical99}.}
\label{OpticalConduct}
\end{figure}

While the high resolution photoemission experiment~\cite{photo2} clearly suggested a metallic ground state with substantial density of states at and around
the Fermi energy, the strongest suggestion in favor of the insulating state has been thought to come from the gap-like structure in the optical conductivity
data~\cite{optical99}, reproduced in Fig.~\ref{OpticalConduct}(a), showing both the intraband contribution, believed to arise from Ni
vacancies, and the interband part, suggestive of an underlying insulating gap of stoichiometric NiS of the order of $\approx 0.2$ eV. In order to make the
comparison between calculated and experimental results on polycrystalline samples~\cite{optical99}, we obtain the  average of the two theoretically computed
inter-band components of the optical conductivity as
$[2\times \sigma_{xx}(\omega) + \sigma_{zz}(\omega)]/3$. The intra-band Drude component of the optical conductivity
is calculated from the bare plasma frequencies (see SM).
We first choose the value
of $U$ to be 3.5 eV, such that the calculated band gap of NiS is 0.2 eV, as claimed in Ref.~\onlinecite{optical99}. The calculated optical conductivity
(Fig.~\ref{OpticalConduct}(a)) is obviously inconsistent
with the experimental result. The reason for this disagreement is quite simple to understand, though this was not appreciated before in spite of the extensive work of several decades on NiS. Associating the sharp drop
in the optical conductivity with a gap in the underlying electronic structure, as interpreted in Ref.~\cite{optical99}, is only possible for direct band gap
materials, while antiferromagnetic NiS has an indirect band gap in the insulating state; we stress here that this is a robust claim and is not dependent on the
specific details of any given calculation. We also show the calculated results with $U$ = 3.0 eV, such that the band gap in the electronic structure
is $\approx$ 0; clearly the calculated optical conductivity  is still significantly shifted towards the higher energy compared to the experimental
data, thereby eliminating any hope to interpret the experimentally observed optical conductivity within an insulating state. A better description of the
experimental results is obtained for $U$ = 2.3 eV (Fig.~\ref{OpticalConduct}(a)), clearly establishing that experimental optical
conductivity is
consistent only with the metallic state, in sharp contrast to the exactly opposite conclusion drawn so far in the literature.
\par
The calculated optical conductivity for $U$~=~2.3~eV in a larger energy window shows a feature-for-feature similarity when compared with the
experimental results except that the calculated features appear systematically at increasingly higher energy.
However a contraction of the energy scale by $\sim$ 30$\%$ brings the calculated result in excellent agreement with the experimental data
as displayed in Fig.~\ref{OpticalConduct}(b).
The necessity for the energy contraction is well-known in the literature whenever calculated spectral features are compared with experimental data
involving unoccupied states, such as Bremsstrahlung Isochromat and X-ray absorption spectroscopies as illustrated in Ref.~\cite{spectra}.
The need for this contraction may arise from limitations of density functional theories in describing the unoccupied states. Another reason can be the neglect of the
electron-hole Coulomb interaction in obtaining the calculated results, while the experimental result is invariably influenced by such interactions.
\par
We have also carried out additional electronic structure calculations for NiS using range of new techniques (e.g. G0W0, scGW0 and scGW)
(see SM)
and find that the essential features of the antiferromagnetic state, discussed above are independent of the calculation scheme adopted, establishing
the validity and the robustness of our conclusion.
While this reinterpretation of the
optical conductivity data in conjunction with the high resolution photoemission results~\cite{photo2} suffice to establish the ground state of the antiferromagnetic
NiS to be metallic, we further require that the present calculation be consistent with other known experimental results from the LT phase of NiS, thereby providing
further credence to the unusual metallic ground state of NiS described here. In doing so, we fix the value of $U$ at 2.3 eV, as suggested by the comparison in Fig.~\ref{OpticalConduct}(a).
\par
\begin{figure}
\includegraphics[width=\columnwidth]{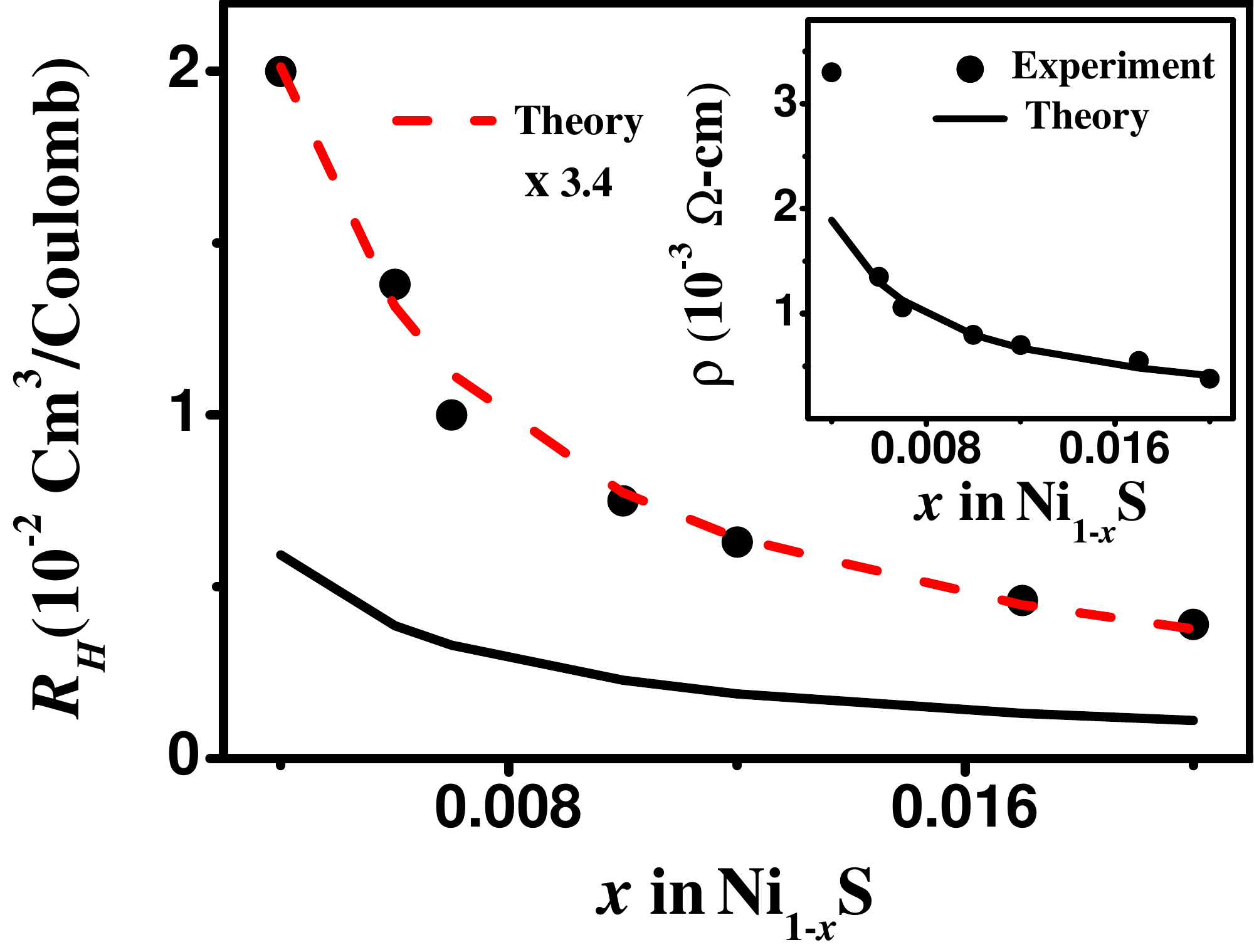}
\caption{Comparison of experimental (filled circles), calculated (solid line) and scaled calculated values (dashed line) for Hall coefficient as a function of Ni
vacancy ($x$). Experimental data are from Ref.~\onlinecite{resistivity}. The inset shows the same for the resistivity.}
\label{transport}
\end{figure}

Based on our results on stoichiometric NiS, we compute the Hall coefficient ($R_H$)  and resistivity($\rho$) in the frame work of Boltzmann transport
theory (details in the SM)
as a function of the Ni vacancy $x$, where the introduction of holes via Ni vacancy is modeled by the rigid band approximation. In addition to the carrier concentration,
we have evaluated effective carrier masses $m^\ast_e$, $m^\ast_{lh}$, and $m^\ast_{hh}$, associated with the electron, light hole and heavy hole carriers.
Fig.~\ref{transport} shows a comparison of the theoretically calculated Hall resistivity ($R_H$) of Ni$_{1-x}$S as a function of $x$
compared with the experimentally
obtained results. If we scale the computed values by a single constant (= 3.4) to match the experimental value at one point, say $x=0.01$,
we find a near perfect match between the calculated  and experimental $R_H(x)$ for all other compositions. Thus, present results are able to describe the dependence of $R_H$ on $x$
exceedingly well with a quantitative agreement well within an order of magnitude of experimental values, thereby establishing a remarkable success of the
description of NiS as a self-doped, nearly compensated antiferromagnetic metal.
\par
We note that the mobility is experimentally obtained as a ratio between the $R_H$ and the resistivity, $\rho$, as in Ref.~\onlinecite{resistivity}. Thus, the scaled, calculated
$R_H$ in conjunction with the experimentally obtained mobility from Ref.~\onlinecite{resistivity} is bound to provide a very good description of the resistivity
as a function of $x$  as well. Going beyond this obvious point, we estimate the mobilities of different charge-carriers at different $x$-values based on our
calculations, as detailed in SM, instead of taking them from experimental results. We note that the mobility is inversely proportional to the effective
mass of a band. We adjust this proportionality constant to equate the computed resistivity with experimentally observed value for $x$ = 0.01, and then use the same
constant to compute the mobility and, therefore, the resistivity of all other compositions without any adjustable parameter. The resulting calculated
resistivity as a function of $x$ is shown in the inset to Fig.~\ref{transport} compared to the experimental data, exhibiting a good match at all
compositions except for the lowest $x$ value. We note that a slightly incorrect estimation of the experimental $x$ value at this composition can easily account for
the discrepancy, since the resistivity rises sharply in this $x$ regime with decreasing $x$.

\par
We now address the experimentally observed nonmagnetic-antiferromagnetic metal-to-metal transition in NiS across $T_c$ = 265 K.
Within our approach, this requires a substantial change in the value of the
intra-atomic Coulomb interaction strength ($U$) in the order of at least 1.5 eV
between the two phases, as evident in Fig.~\ref{bandgap}.
In order to understand such a substantial change in the strength of the
intra-atomic Coulomb interaction  between the two states, we have computed
the value of $U$ self-consistently within the constrained random-phase
approximation \cite{Spex,cRPA}. For the HT crystal structure we obtain $U$ = 1.8 eV
and for the LT antiferromagnetic state $U$ = 3.7 eV. While these values by themselves would
suggest incorrect electronic structures for both phases according to Fig.~\ref{bandgap},
for example a magnetic state for the HT structure, we note the known
underestimation of $U$ values obtained by requiring the right
ground state properties within LSDA+U compared to the one obtained from such
constrained approaches~\cite{NiO_prb2000}. Instead it is important to note here that significantly, we find an appreciable
increase of about 1.9 eV in the calculated strength of $U$ in going from the
nonmagnetic to the antiferromagnetic state, as also suggested by our analysis
of the experimental data in terms of LSDA+U approach presented in this work.
This increase in $U$ in going from the nonmagnetic state to the
antiferromagnetic state can be easily understood in terms
of its dependence on the magnetic moment. In antiferromagnets, a shift of the 
spectral-weight around the Fermi level reduces the DOS, which leads to
smaller polarization and thus, to larger $U$ values. Larger the magnetic moment
stronger becomes the Coulomb interaction strength, $U$.
\begin{figure}
\includegraphics[width=\columnwidth]{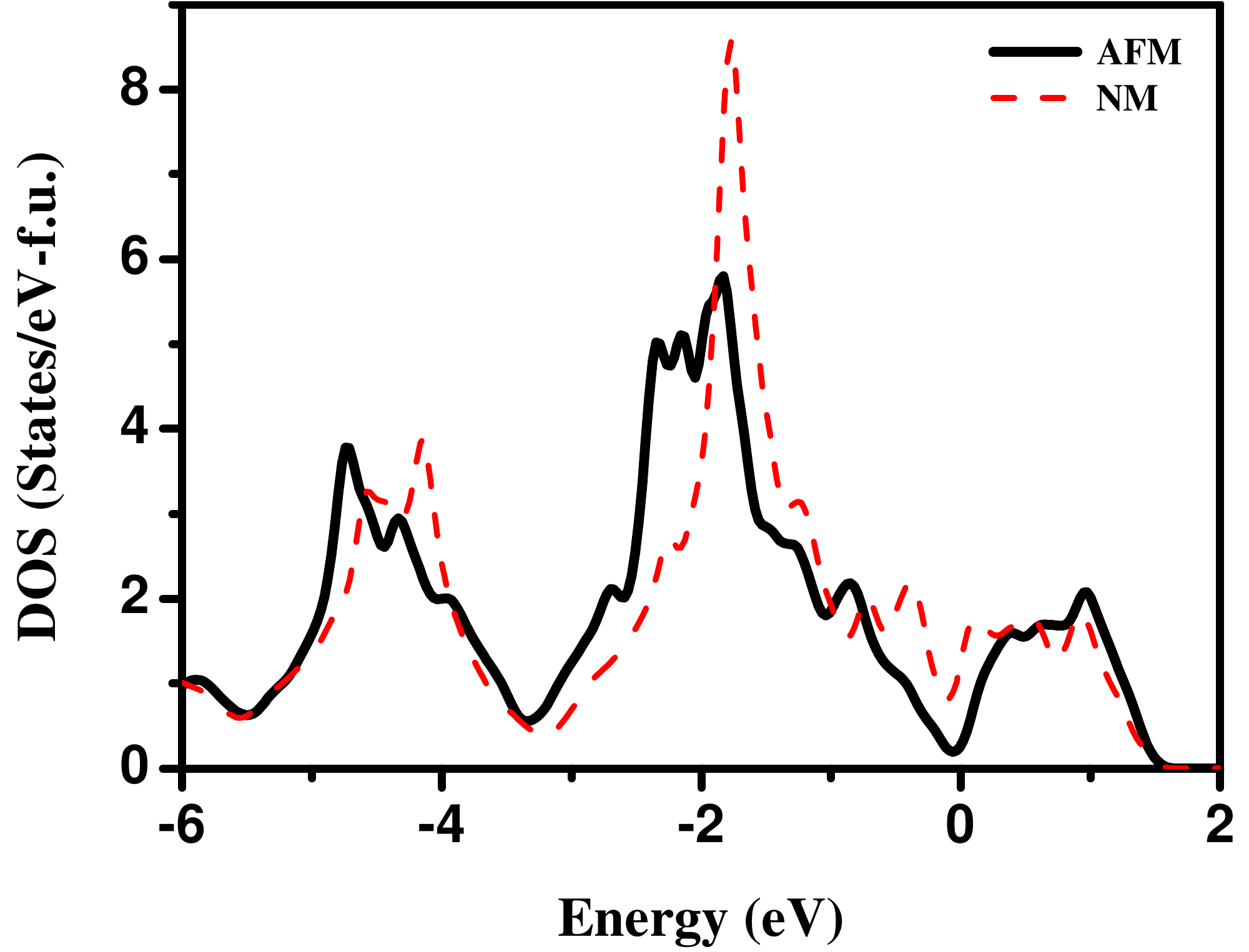}
\caption{Total DOS corresponding to the AFM and NM phase for $U$ = 2.3 eV. The energy scale is referenced to the Fermi energy.}
\label{dos}
\end{figure}
\par
Finally we show the density of states obtained for the self-doped, antiferromagnetic metallic phase
with $U$ = 2.3 eV and that for the nonmagnetic metallic phase in Fig.~\ref{dos}.
While the two DOS are roughly similar, we can clearly see two distinctive changes that should be visible in photoemission experiments.
First is the small movement of the most intense part of the DOS near 2 eV towards a higher binding energy in the self-doped, antiferromagnetic metallic state.
This is in agreement with a similar shift of the main photoemission peak reported for the LT phase of NiS in Ref.~\onlinecite{photo1}. The other visible change is
a depletion of the DOS at $E_F$ for the self-doped antiferromagnetic state compared to the nonmagnetic one.
This is exactly what has been reported in Ref.~\onlinecite{photo2}, with experimental photoemission data showing a decrease in the spectral weight at $E_F$,
though it remains finite, at the lowest temperature. The decrease in the DOS is related to a decrease in the Fermi volume in the LT
phase, associated with opening of gaps at the Fermi energy along several momentum directions.
\section*{Discussion}
We have shown that the LT phase of NiS in the NiAs structure has to be described as a novel self-doped, nearly compensated antiferromagnetic
metal close to the $p$-$d$ metal/covalent insulator phase line. A host of physical properties calculated on the basis of this description of the ground state of NiS is in remarkable
agreement with all experimentally determined properties of NiS, establishing the much discussed temperature driven phase transition of NiS at about 265 K to be one
between a Pauli paramagnetic metal and this unusual self doped antiferromagnetic metal. We have also shown that the transition from the HT nonmagnetic metallic state to the LT antiferromagnetic unusual metal state is driven by a substantial increase in the
intra-atomic Coulomb interaction strength, that remains sub-critical to drive the system in to an insulating state.
\section*{Method}
All  calculations reported in this work are carried out using full potential linearized augmented plane wave (FPLAPW) plus local orbitals method as embodied in WIEN2k
code~\cite{wien2k}.
The muffin-tin radii ($R_{MT}$) of Ni and S are chosen to be 1.23~\r{A}. and 1.09~\r{A}, respectively. To achieve energy convergence of the eigenvalues,
the wave functions
in the interstitial region were expanded in plane waves with a cutoff $R_{MT}k_{max}$=7, where $R_{MT}$ denotes the smallest atomic sphere radius and $k_{max}$
represents the magnitude
of the largest k vector in the plane wave expansion. The valence wave functions inside the spheres are expanded up to $l_{max}$=10, while the charge density was Fourier
expanded up to $G_{max}$=12. The Brillouin-zone integration was  performed with a modified tetrahedron method using 394 and 2839 special k points in
the irreducible
part of the Brillouin-zone to achieve self-consistency and for optical calculations, respectively.
Exchange and correlation effects are treated within the DFT using LSDA~\cite{LSDAPARA} and
a Hubbard $U$ was included in the framework~\cite{AMF} of LSDA+U for the description of electron
correlations. Hexagonal NiS crystallizes in the NiAs structure~\cite{cryststruct}.
Since we are interested in the ground state of NiS, all results reported here are for the LT structure, unless otherwise stated.

\section*{Acknowledgments}
The authors thank Department of Science and Technology as well as Council of Scientific and Industrial Research, Government of India for financial support. Discussions with  C. Friedrich are gratefully acknowledged. This work has been supported by the DFG through the Research Unit FOR-1346.
\section*{Author contributions}
DDS suggested the problem. SP carried out all calculations based on WIEN2k and VASP under the supervision of ID. The results were discussed and interpreted by DDS, ID, and SP. Calculations of $U$ within constrained RPA were carried out by ES and SB. The manuscript was largely written by DDS with inputs from all authors and revised according to suggestions received from all.

\end{document}


\maketitle

\section{Partial DOS, Orbital projected band structure (Fatband) and Fermi Surface}
\begin{figure}[h]
\begin{center}
\includegraphics[scale=0.32]{supp1.eps}
\caption{Partial densities of states of Ni-d and S-p for both the spin channel.}
\end{center}
\end{figure}

\begin{figure}[h]
\begin{center}
\includegraphics[scale=0.6]{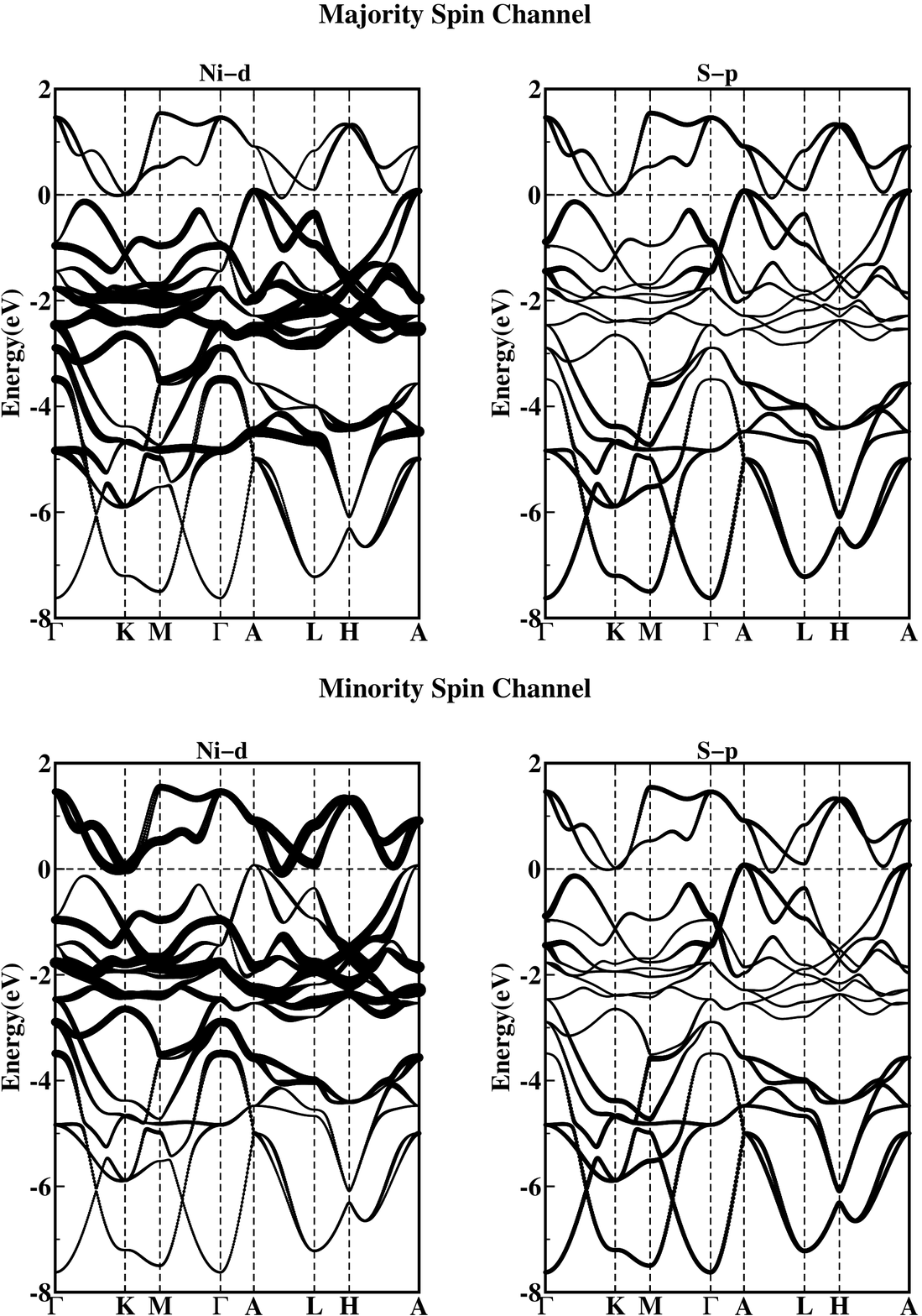}
\caption{Fatband representation of the Ni-d and S-p states for both the spin channel.}
\end{center}
\end{figure}

\begin{figure}[h]
\begin{center}
\includegraphics[scale=0.4]{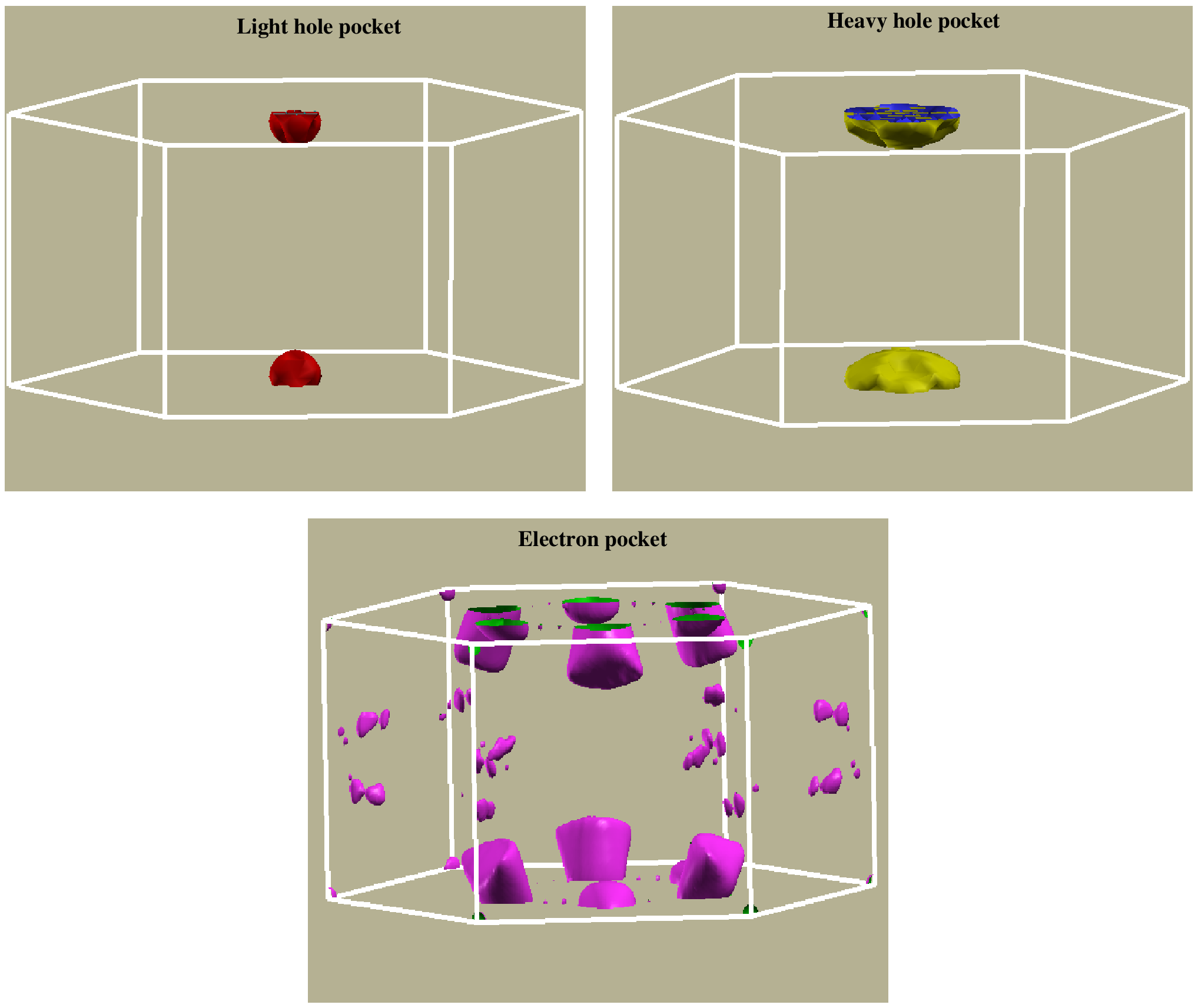}
\caption{Fermi surface plot corresponding to the electron and hole pockets. There are two hole pockets around the A (0, 0, $\frac{1}{2}$) point.
Electron pockets are seen around K ($\frac{1}{3}$, $\frac{1}{3}$, 0)  point and some other points in the BZ. The coordinates given in the parenthesis, are
in terms of reciprocal lattice vectors.}
\end{center}
\end{figure}
\clearpage
\section{Hall coefficient and resistivity}
The Hall coefficient has been evaluated using the following expression
\begin{equation}
R_H=\frac{1}{e} \frac{n_{lh}\mu_{lh}^2+n_{hh}\mu_{hh}^2-n_e\mu_e^2}{(n_{lh}\mu_{lh}+n_{hh}\mu_{hh}+n_e\mu_e)^2}  \nonumber
\end{equation}
where $n_{lh}$, $n_{hh}$, $n_e$ are the carrier concentrations of light hole, heavy hole and electron respectively. $\mu_{lh}$, $\mu_{hh}$ and $\mu_e$ are the
corresponding mobilities.\\
Since mobility is inversely proportional to the effective mass, the above expression can also be written in the following way
\begin{equation}
R_H=\frac{1}{e} \frac{\frac{n_{lh}}{m_{lh}^{*2}}+\frac{n_{hh}}{m_{hh}^{*2}}-\frac{n_e}{m_e^{*2}}}{(\frac{n_{lh}}{m_{lh}^*}+\frac{n_{hh}}{m_{hh}^*}+\frac{n_e}{m_e^*})^2}  \nonumber
\end{equation}
where $m_{lh}^*$, $m_{hh}^*$, $n_e^*$ are the effective masses of light hole, heavy hole and electron respectively.\\
To compute the resistivity, the following expression has been used
\begin{equation}
 \rho=\frac{1}{e(n_{lh}\mu_{lh} + n_{hh}\mu_{hh} + n_{e}\mu_{e})} \nonumber
\end{equation}
If we assume that the proportionality constant between the effective mass and mobility as $K$, the above expression can be written as
\begin{equation}
 \rho=\frac{1}{eK(\frac{n_{lh}}{m_{lh}^*}+\frac{n_{hh}}{m_{hh}^*}+\frac{n_e}{m_e^*})} \nonumber
\end{equation}
We evaluate $K$ by requiring that the calculated resistivity be identical to the experimental resistivity at $x=0.01$. This estimated value of $K$ is used to determine the
resistivity for all the other compounds with different $x$-values.

\section{Optical Conductivity}
The optical properties of a material is described by the dielectric function $\varepsilon(\omega) = \varepsilon^r(\omega) + i\varepsilon^i(\omega)$. Interband contribution
to the imaginary part
of $\varepsilon(\omega)$ is calculated by convoluting the joint density of states with the dipole matrix elements between the occupied and the unoccupied states.
We use the following expression to obtain the imaginary part of the dielectric tensor. 
\begin{equation}
\varepsilon^i_{\alpha\beta}(\omega)=\frac{4\pi e^2}{m^2\omega^2}\sum_{c,v}\int dk \langle\psi_{c,k}|p^\alpha|\psi_{v,k} \rangle \langle\psi_{v,k}|p^\beta|\psi_{c,k} \rangle \delta(\epsilon_{c,k}-\epsilon_{v,k}-\omega) \nonumber
\end{equation}
where e is the electronic charge, m is the mass of the electron and $\omega$ is the energy of the incident photon.
$\psi_{c,k}$, $\psi_{v,k}$ are the conduction band and valence band wavefunction at point $k$ and $\epsilon_{c,k}$, $\epsilon_{v,k}$ are the corresponding energy eigenvalues. \\
Real part of the optical conductivity is given by
\begin{equation}
\sigma_{\alpha\beta}(\omega) = \frac{\omega}{4\pi} \varepsilon^i_{\alpha\beta}(\omega) \nonumber
\end{equation}
For the hexagonal system, due to the symmetry, all the off-diagonal components of the optical conductivity are zero and diagonal components obey the
relationship: $\sigma_{xx}=\sigma_{yy}\neq\sigma_{zz}$.
The interband part of the optical conductivity is broadened by a Lorentzian function. The broadening parameter of the Lorentzian ($\Gamma$) is taken to be
linearly dependent with the photon energy, having the form: $\Gamma =0.03 + 0.05 \omega$. 
\par
The intraband part of the optical conductivity (Drude term) is obtained from plasma frequency ($\omega_p$). To compute the plasma frequency, we use the following
expression
\begin{equation}
\omega_{p,\alpha\alpha}^2 = \frac{e^2}{m^2\pi^2}\sum_l\int dk |\langle\psi_{l,k}|p^\alpha|\psi_{l,k}\rangle|^2\delta(\epsilon_{l,k}-\epsilon_F) \nonumber
\end{equation}
where $\epsilon_F$ is the Fermi energy.
Intraband component of the real part of the optical conductivity is given by
\begin{equation}
\sigma_{\alpha\alpha}(\omega)=\frac{\omega_{p,\alpha\alpha}^2}{4\pi} \frac{\varGamma}{\omega^2+\varGamma^2} \nonumber
\end{equation}
where $\varGamma$ is the broadening parameter for the Drude term.
\par
Finally we add the intraband and interband part of each component of the optical conductivity and take an average of all the components to compare with the experimental
optical conductivity.

\section{Various forms of GW calculations}
\begin{figure}
\begin{center}
\includegraphics[scale=0.45]{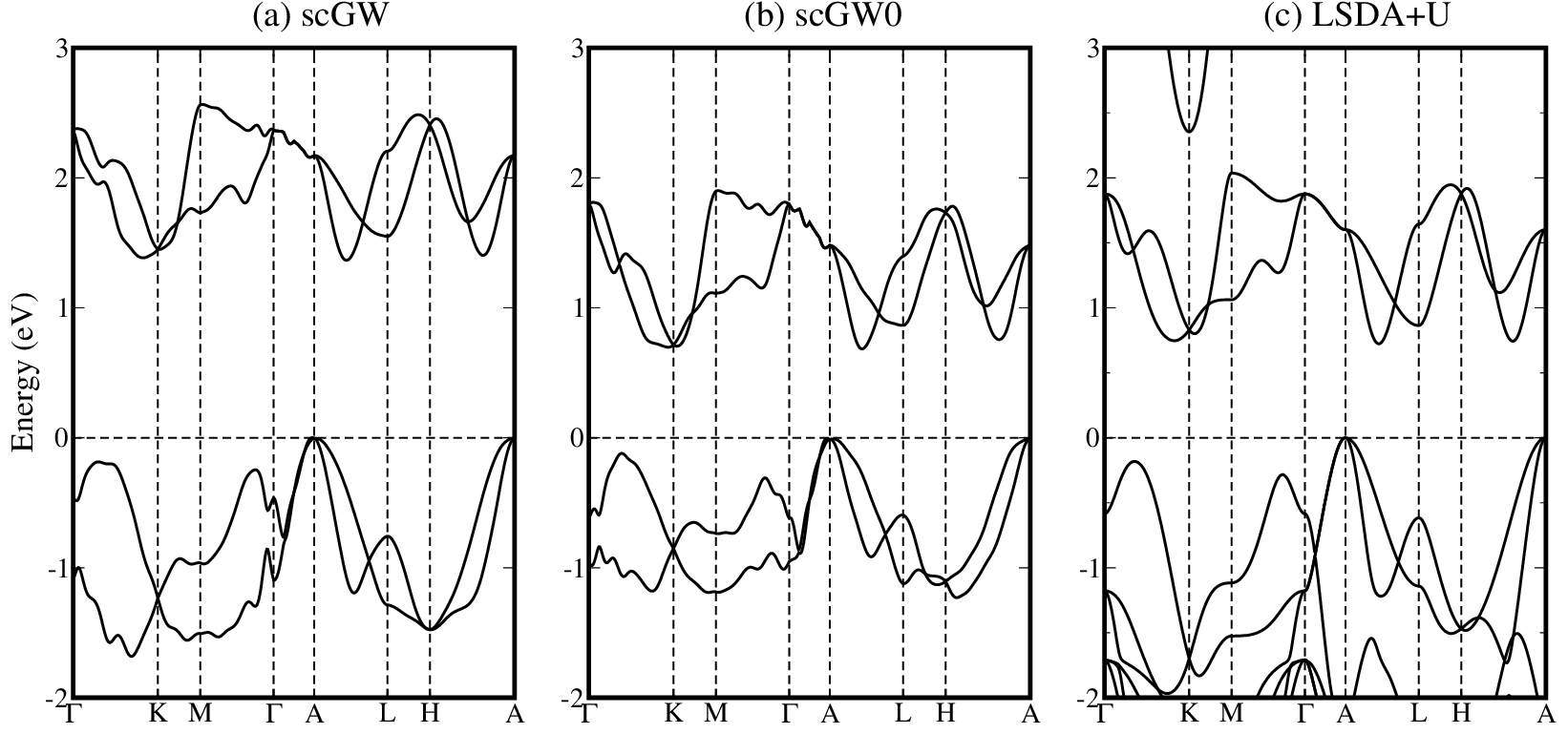}
\caption{Band dispersion of NiS in the antiferromagnetic insulating regime plotted along various high symmetry directions computed within
(a) scGW, (b) scGW0, and (c) LSDA+U (with $U$ = 5.1 eV) methods. The figure shows that the essential qualitative parts of the electronic band dispersions
remain identical, irrespective of the calculation method used, indicating NiS to be an indirect gap insulator.}
\label{scGW_bands}
\end{center}
\end{figure}

We have carried out the various forms of GW calculations as implemented in the VASP code~\cite{vasp1,vasp2} to show that all important features of the electronic structure of NiS are independent
of the specific calculation scheme adopted. In Fig.~\ref{scGW_bands} we have displayed the band dispersion for the low temperature phase of NiS
along various high symmetry directions calculated within (a) scGW, (b) scGW0 and (c) LSDA+U (with $U$ = 5.1 eV) approach.
The scGW calculations (Fig.~\ref{scGW_bands}-(a)) shows a wide ($~$1.36 eV) indirect gap semiconductor/insulator. The scGW0 calculation
yields a significantly reduced gap of 0.68 eV (see Fig.~\ref{scGW_bands}-(b)) but all the essential features of band dispersions remain the same as in the case of scGW
results (Fig.~\ref{scGW_bands}-(a)). The LSDA+U calculations with $U$ = 5.1 eV, shown in Fig.~\ref{scGW_bands}-(c) gives 
essentially the same band dispersions as in scGW0 with an identical band gap of 0.68 eV. We can also obtain a bandgap of 1.36 eV within LSDA+U with a $U$ of 7.1 eV.
Thus, in the antiferromagnetic insulating regime, the essential qualitative parts of the electronic band dispersions remain identical, irrespective of the calculation
method (scGW, scGW0 or LSDA+U), indicating an indirect gap insulator with very similar dispersions. Therefore, all results, computed on the basis of these
results also remain the same; specifically, the experimental optical data, shown as Fig. 3 of our manuscript, remain entirely incompatible with all three results
shown in Fig.~\ref{scGW_bands}, as the smallest direct gap, defining the smallest energy for optical absorption,
are 1.75, 0.98, and 1.09 eV for scGW, scGW0 and LSDA+U respectively, whereas the experimental value
is 0.2 eV. Therefore, it is clear that both scGW and scGW0 are not suitable to describe the low temperature phase of NiS. 
\par
\begin{figure}
\begin{center}
\includegraphics[scale=0.45]{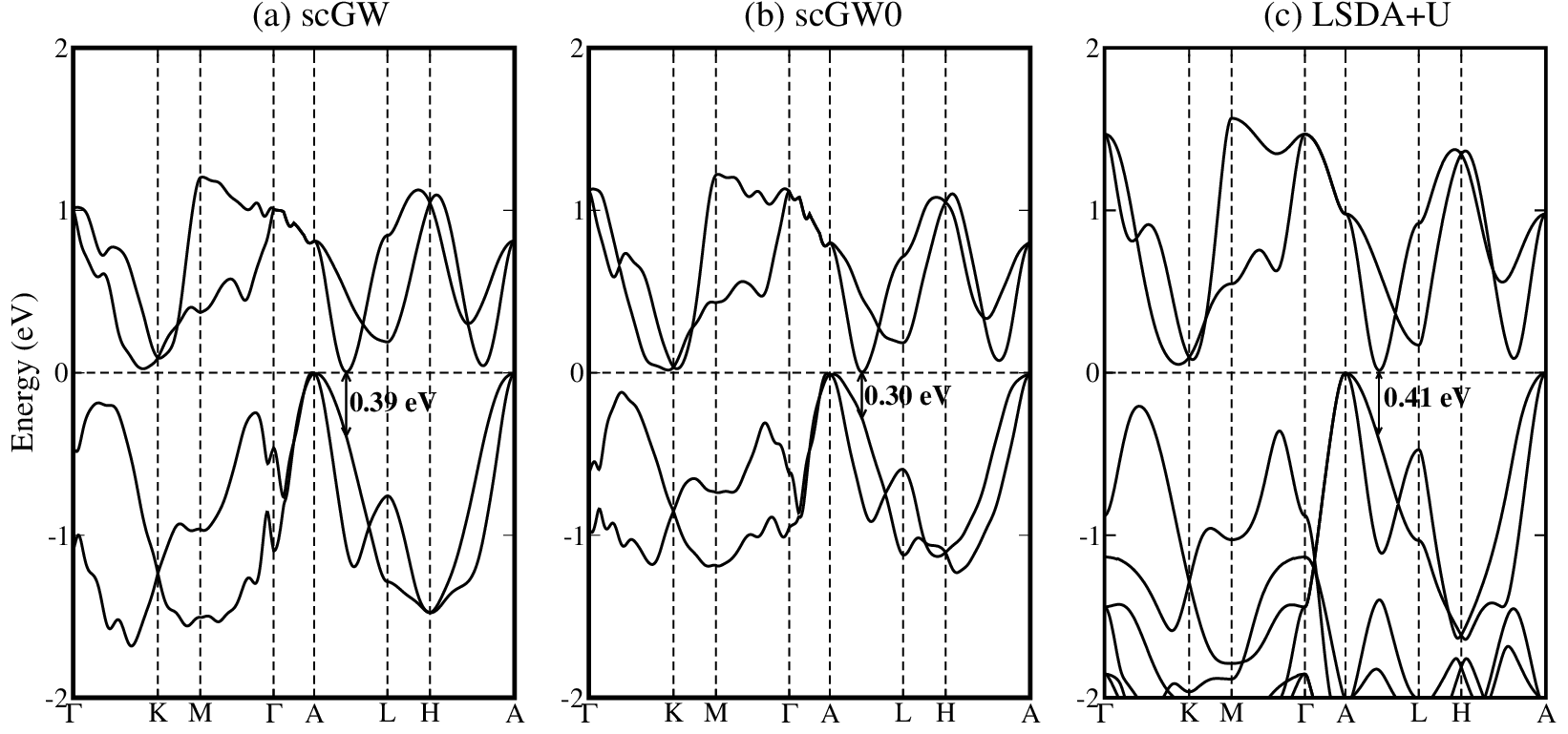}
\caption{Band dispersion for AFM NiS obtained by (a) scGW and (b) scGW0, where the band gap is made infinitesimally small using scissors operator.
(c) Band dispersion computed within LSDA+U method with $U$ = 3.0 eV which gives an infinitesimally small band gap. Arrow in each panel indicates the optical gap
where in each case it is much larger than the experimental gap of 0.2 eV.}
\label{scGW_scissor}
\end{center}
\end{figure}

Next we have performed the scissors operation on the scGW0 and the scGW band dispersions by vertically moving the empty
conduction band down towards the occupied valence band to decrease both the band gap and the optical gap continuously. If we do this to make the bandgap
infinitesimally small, we are still left with an optical (direct) gap of 0.39 and 0.30 eV for the scGW and scGW0, respectively, as shown
in Fig.~\ref{scGW_scissor}-(a) and ~\ref{scGW_scissor}-(b), once again totally inconsistent with experimental finding of a 0.2 eV optical gap.
Thus, the qualitative features of band dispersions obtained with scGW and scGW0 already establish that the only way calculated results can be consistent with
the experimental results is to push the conduction band further down by about $~$0.2-0.3 eV, making the system metallic,
resulting from an indirect overlap of the valence and the conduction bands of these dispersions, independent of the calculation method employed,
such that the direct gap can be in the order of 0.2 eV required by the experimental result.
\par
Avoiding this arbitrary reduction of the bandgap, we can alternately use LSDA+U approach where by tuning $U$ as a parameter, we can reduce the bandgap,
exploring whether any self-consistent gapped or insulating band dispersion can be made to agree with the experimental data. Fig. 3 in the manuscript
shows that this is not possible, since even an infinitesimally small bandgap (obtained for $U$ = 3.0 eV), being an indirect one, has a direct (or optical) gap
of $~$0.41 eV (see Fig.~\ref{scGW_scissor}-(c)) which is twice as large as what is seen in the experiment, (see Fig. 3 in the manuscript).
This shows that independent of the method of calculation, experimental optical data requires the system to be ungapped or metallic, with an indirect overlap
of the top of the valence band at the A-point and the bottom of the conduction band at about half-way along the A-L direction.
\par
\begin{figure}
\begin{center}
\includegraphics[scale=0.45]{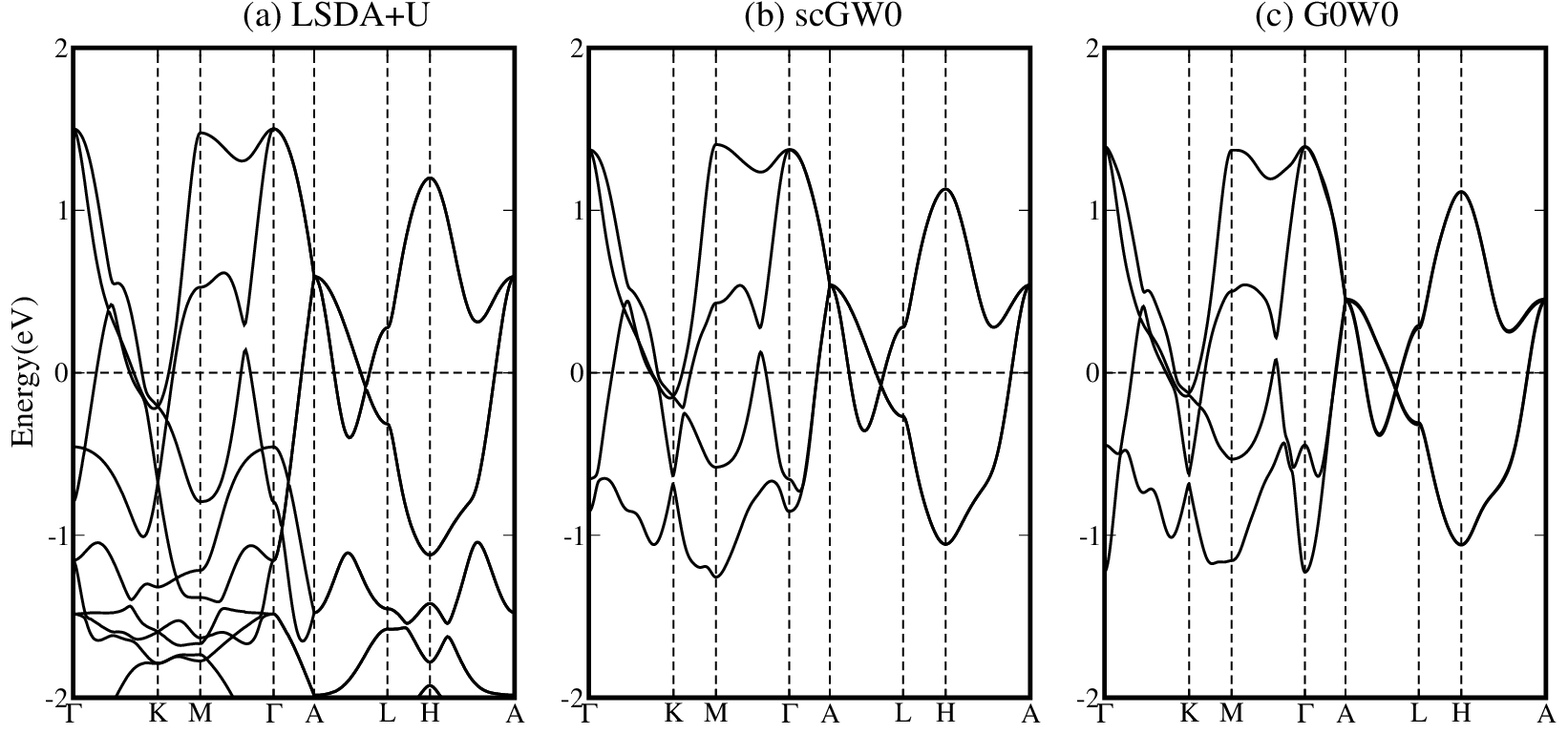}
\caption{Band dispersion along various high symmetry directions computed within (a) LSDA+U (with $U$ = 0.8 eV), (b) scGW0 and (c) G0W0 (on PBE state) approaches
for the high temperature Pauli paramagnetic, metallic phase of NiS.}
\label{GW_bands_HT}
\end{center}
\end{figure}

In Fig.~\ref{GW_bands_HT}, we show electron energy dispersions obtained from LDA+U (with $U$ = 0.8 eV) (Fig.~\ref{GW_bands_HT}-(a)), scGW0 (Fig.~\ref{GW_bands_HT}-(b))
and G0W0 on PBE (Fig.~\ref{GW_bands_HT}-(c))
for the high temperature Pauli paramagnetic, metallic phase. The comparison in Fig.~\ref{GW_bands_HT} makes it clear that the three approaches give
essentially the same results, once again making our conclusions concerning the high temperature nonmagnetic, metallic state independent of the calculation 
technique employed.
\par
\begin{figure}
\begin{center}
\includegraphics[scale=0.5]{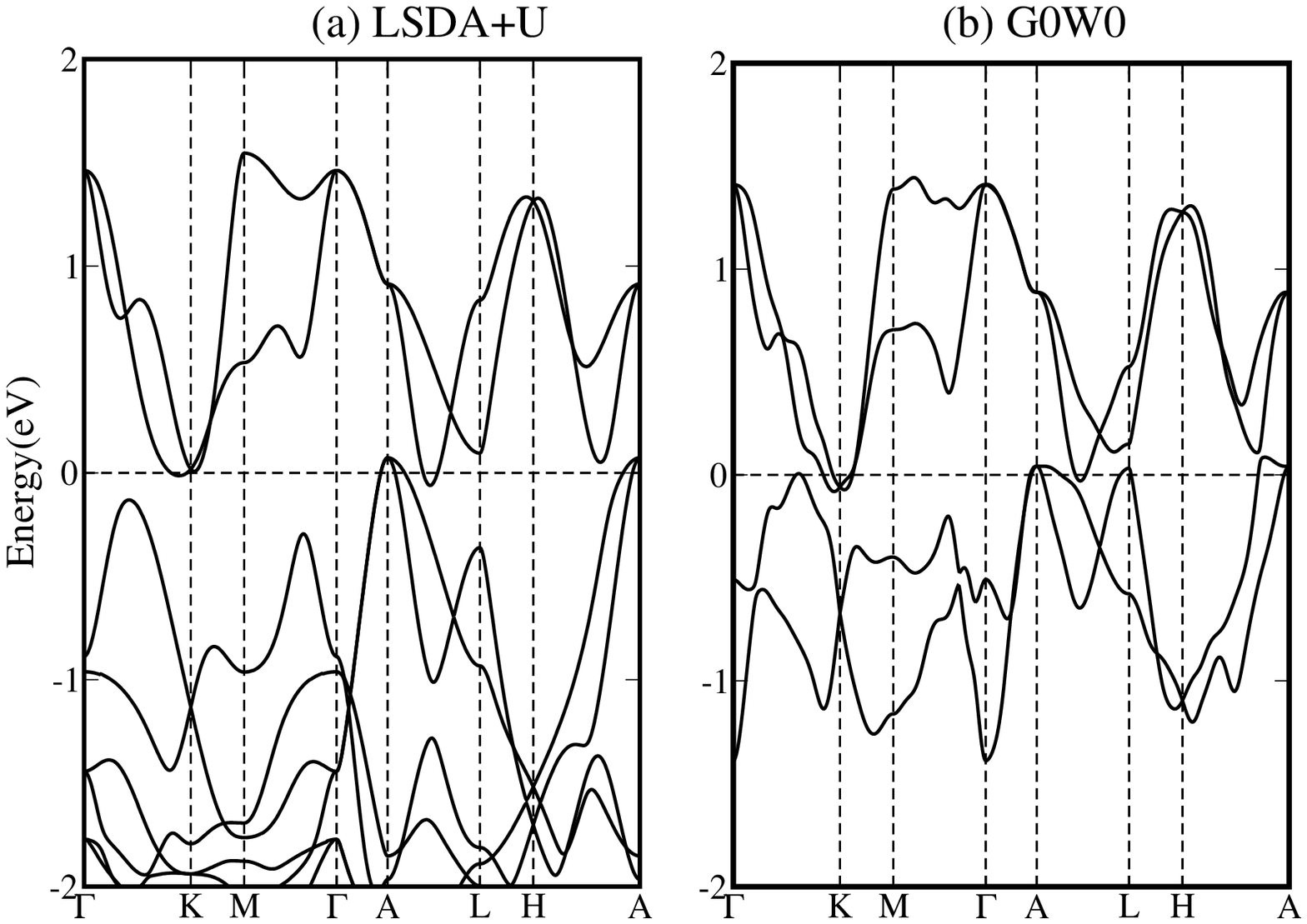}
\caption{Band dispersion of NiS along various high symmetry directions computed within (a) LSDA+U (with $U$ = 2.3 eV), and (b) G0W0 on PBE approaches
for the low temperature antiferromagnetic metallic state. All the calculations capture the important features of the band dispersions, namely the self-doping and
compensation as discussed in the text.}
\label{GW_bands_LT}
\end{center}
\end{figure}

Finally we present the
energy dispersions for the antiferromagnetic metallic state within LSDA+U (with $U$ = 2.3 eV) (Fig.~\ref{GW_bands_LT}-(a)) and G0W0 on PBE (Fig.~\ref{GW_bands_LT}-(b)).
Clearly, once again, description of energy dispersions from these two techniques are almost the same, with all important features,
discussed in our manuscript, namely self-doping and compensation, being realized in every case. This is a very strong evidence that our novel interpretation
of the unusual, low temperature phase of NiS in terms of a novel self-doped, nearly compensated, antiferromagnetic, low-density metal is robust with respect
to the specific calculation scheme adopted by us.
\par
\begin{figure}
\centering
\includegraphics[scale=0.45]{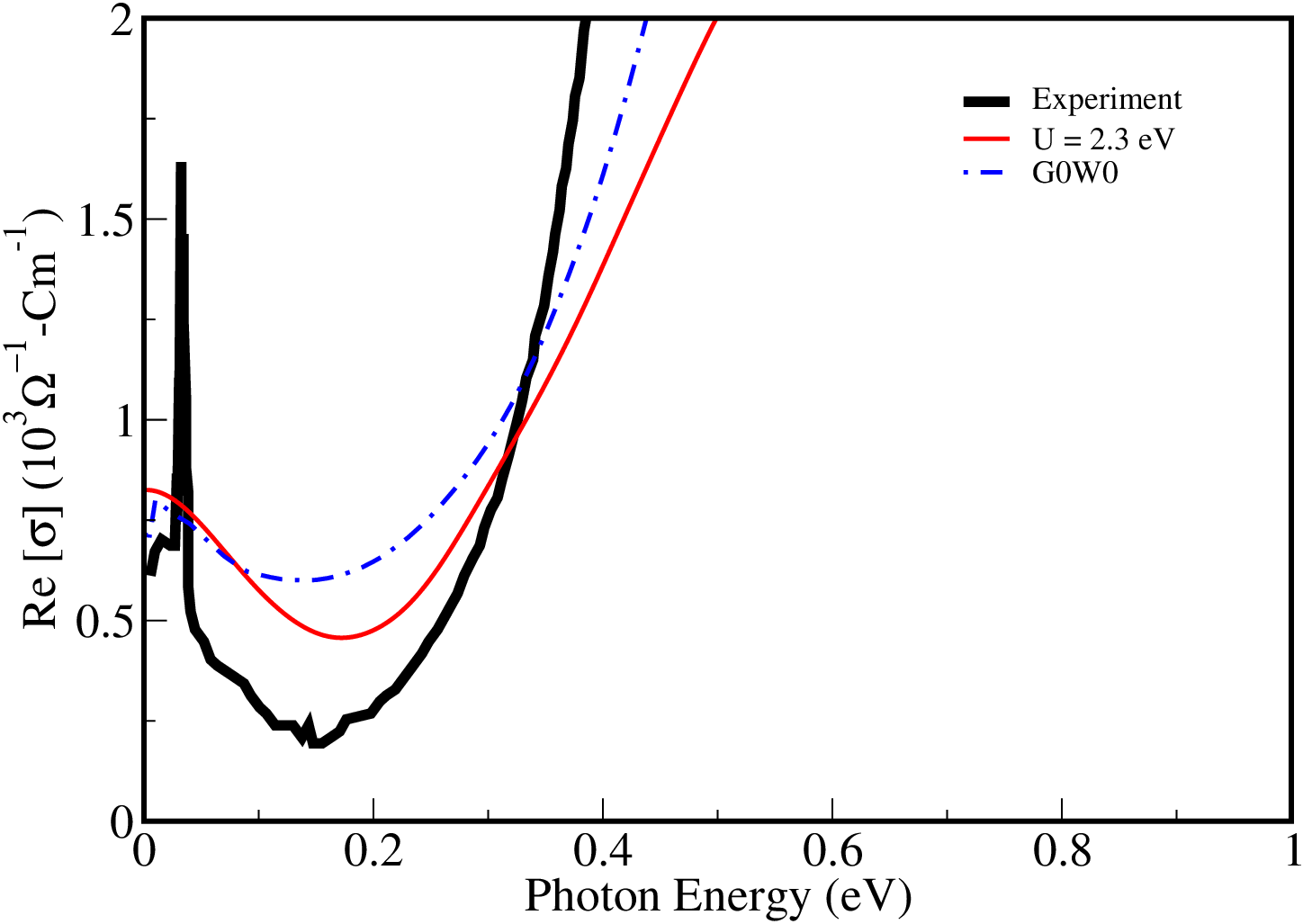}
\caption{A comparison of the G0W0 optical conductivity with the experimental data and the calculated LSDA+U results (with $U$ = 2.3 eV). The good agreement of
G0W0 with the experimental results and LSDA+U approach establish the robustness of our conclusion as discussed in the text.}
\label{optics_gw_lsdau}
\end{figure}

While we do not have the possibility of calculating the matrix elements relevant for the optical conductivity data within any of the GW approaches, we note
that the joint density of states, when weighted by the matrix elements, yields the optical data. Thus, one expects that the qualitative features of the optical data
to be reasonably well captured by the energy distribution of the joint density of states. So we have computed the joint density of sates based on the electronic
structure results from the G0W0 on PBE calculation of the low temperature crystal structure of NiS; the result is compared with the experimental data along with
the calculated LSDA+U results in Fig.~\ref{optics_gw_lsdau}. Thus, the G0W0 result, without any adjustable parameter and yielding the self-doped metallic state,
provides a good description of the experimental data, as shown in Fig.~\ref{optics_gw_lsdau}, and corresponds closely to the calculated result from
LSDA+U approach (which we use for all results in our manuscript), thereby once again establishing the only possible description of the
ground state of the low temperature NiS is in terms of the novel state proposed by us and this conclusion is truly independent of the calculation methodology. 
\linebreak
\linebreak
\linebreak
\linebreak


